\documentclass[letterpaper, 10pt, conference]{ieeeconf}

\usepackage{cite}
\usepackage{amsmath,amssymb,amsfonts,latexsym,mathrsfs}
\usepackage{nccmath}
\usepackage{bm}
\usepackage{url}
\usepackage{subfig}
\usepackage{algorithmic}
\usepackage{graphicx}
\usepackage{textcomp}
\usepackage{comment}
\usepackage[switch]{lineno}
\usepackage{todonotes}
\IEEEoverridecommandlockouts
\graphicspath{{figures/}}

\newcommand{\norm}[1]{\|#1\|}
\renewcommand{\vec}{\bm}

\newcommand{\Tr}{\operatorname{Tr}}

\newcommand{\A}{\bm A}

\renewcommand{\S}{\bm S}
\newcommand{\I}{\bm I}

\newcommand{\bTheta}{\bm \Theta}
\newcommand{\T}{\mathbf{T}}
\renewcommand{\H}{\mathbf{H}}

\begin{document}

\title{Applying classical control techniques to quantum systems: entanglement versus stability margin and other limitations}
\author{C.~A.~Weidner \and
S.~G.~Schirmer \and 
F.~C.~Langbein \and 
E.~Jonckheere
\thanks{CAW is with the Quantum Engineering Technology Laboratories, H. H. Wills Physics Laboratory and Department of Electrical and Electronic Engineering, University of Bristol, Bristol BS8 1FD, United Kingdom (e-mail: c.weidner@bristol.ac.uk).}
\thanks{SGS is with the Faculty of Science \& Engineering, Swansea University, Swansea SA2 8PP, UK (e-mail: s.m.shermer@gmail.com).}
\thanks{FCL is with the School of Computer Science and Informatics, Cardiff University, Cardiff CF24 4AG, UK (e-mail: frank@langbein.org).}
\thanks{EAJ is with the Department of Electrical and Computer Engineering, University of Southern California, Los Angeles, CA 90089 USA (e-mail: jonckhee@usc.edu).}
}
\maketitle
\thispagestyle{empty}
\pagestyle{empty}

\begin{abstract}
Development of robust quantum control has been challenging and there are numerous obstacles to applying classical robust control to quantum system including bilinearity, marginal stability, state preparation errors, nonlinear figures of merit.  The requirement of marginal stability, while not satisfied for closed quantum systems, can be satisfied for open quantum systems where Lindbladian behavior leads to 
non-unitary evolution, and allows for nonzero classical stability margins, but it remains difficult to extract physical insight when classical robust control tools are applied to these systems. We consider a straightforward example of the entanglement between two qubits dissipatively coupled to a lossy cavity and analyze it using the classical stability margin and structured perturbations.  We attempt, where possible, to extract physical insight from these analyses.  Our aim is to highlight where classical robust control can assist in the analysis of quantum systems and identify areas where more work needs to be done to develop specific methods for quantum robust control.
\end{abstract}

\section{\label{sec:intro}Introduction}

Despite the extensive success and ongoing development of robust control theory~\cite{Zhou}, the tools developed were found not to be readily applicable to classical systems~\cite{safonov_lightly_damped}, and they are even less readily adaptable to quantum mechanical systems.  This is increasingly problematic amidst the second quantum revolution, when quantum mechanical devices for computing, networking, sensing, and simulation are moving out of laboratories and into commercial markets~\cite{deutsch_quantum}, where robust control of these devices is critical to their utility in real-world settings~\cite{Petersen2013}. These issues arise because quantum systems can be difficult to cast as linear, time-invariant control systems subject to feedback control stabilization.  Rather, quantum control systems, especially those that evolve unitarily in so-called \emph{closed} systems, are typically open-loop bilinear systems, although several types of feedback control for closed-loop quantum systems, using both measurement-based and coherent feedback, have been developed~\cite{wiseman,Petersen2013}.

Uncertainties in quantum systems can take the form of structured perturbations in state parameters (similar to classical structured uncertainties), but these uncertainties must arise in such a way that the evolution of the system remains physical and follows the law of quantum mechanics (i.e. is Hermitian), so care must be taken when applying the methods of structured uncertainties.  Also, classical robust control is usually not concerned with initial state errors, which mainly affect the transient dynamics, while such errors are considerably more important for quantum systems.

There are perturbations of quantum systems that do not make sense classically at all, such as decoherence.  Other challenges that arise are \emph{degeneracies} that often give rise to multiple poles in a region. Some of these can be easily broken via additional parameters (e.g. electric or magnetic fields)~\cite{von_neumann_wigner}, while others are the result of symmetries and physical constraints (e.g., the requirement that the trace of a density matrix be unity) that cannot be eliminated. In addition, the existence of uncertainty relations (e.g., the inability for one to precisely measure both position and momentum simultaneously) is a uniquely quantum phenomenon that limits the information that one can glean about the state of the system. Finally, while some performance measures such as state transfer or gate fidelities can be expressed as linear functions in the quantum state, other performance measures that are intrinsically quantum, such as state entanglement or squeezing, are inherently nonlinear functions of the quantum state, which poses additional challenges when developing quantum-amenable robust control theories.

Robustness for quantum systems is therefore typically assessed via Monte-Carlo simulations, wherein one samples the parameter space of uncertainties and determines averages and distributions of relevant performance measures~\cite{statistical_control}, although there exist other methods for determining uncertainty~\cite{MogensFelixCarrie}.  Monte-Carlo approaches have their merits, but they are often computationally expensive and rely solely on numerical and statistical analyses.  Other robustness measures, such as the log-sensitivity of the performance measure, are sometimes (but not always) analytically tractable, but they typically only apply in the regime of small perturbations (known as the \emph{perturbative regime} in the quantum literature). Thus it is of interest to develop quantum analogues to tools like mu-analysis and robust performance. 
that can handle perturbations regardless of their size.  However, due to the challenges listed above, this is not straightforward.

Closed quantum systems are, by definition, marginally stable in that all of their poles lie along the imaginary axis.  Instead, \emph{open} quantum systems, that are subject to decoherence or dissipation (i.e., exhibit non-unitary evolution), are classically stable due to the fact that the dissipative effects push the poles into the left half of the plane.

As a result, we can apply classical robust control techniques to open systems~\cite{robust_performance_open}, 
but this comes with an inevitable trade-off: the more non-unitary behavior one includes in a quantum system, the less inherently quantum it becomes in general, in that one can model decoherence and dissipation as the loss of quantum information to the surrounding environment. 

However, one area of quantum control where dissipation is often used to facilitate the creation of entanglement is quantum reservoir engineering. In this paper, we will explore one such system, especially a system of two entangled qubits coupled to one another via a lossy cavity~\cite{motzoi}. Thus, the aim of this paper is to revisit a canonical problem of generation and stabilization of entanglement, formulated in a way that is amenable to control, and compare different performance and robustness measures from simple stability margin analysis to the analysis of structured perturbations. Where possible, we will try and link physical insight to the results, highlighting open questions where they arise, i.e., the limitations of our methods that highlight the need for a more general framework of robust control that is applicable to quantum systems.

\section{\label{sec:problem}Two qubits coupled to a lossy cavity}

\noindent  We begin with the Lindblad open quantum system model
\begin{equation}\label{e:QME1}
  \frac{d}{dt}\rho(t)=-\imath [H,\rho(t)]
  +\sum_{k}\gamma^2_{k} \mathfrak{L}(V_k)\rho(t), \quad \rho(0)=\rho_0,
\end{equation}
where $[A,B] = AB-BA$ represents the commutator, $\rho$ is the density operator over the Hilbert space $\H$ of dimension $N$, $H$ is the Hamiltonian, and the Lindblad superoperator is
\begin{equation}\label{e:QME2}
  \mathfrak{L}(V_k) \rho
  = \sum_k V_k \rho V_k^\dag -\tfrac{1}{2} (V_k^\dag V_k \rho+\rho V_k^\dag V_k),
\end{equation}
where the $V_k$s are the quantum jump operators that are typically problem-specific. We apply this formalism to the combined qubit-cavity system described in Ref.~\cite{motzoi} wherein, upon adiabatic elimination of the cavity via a unitary transformation, the system can be described via Eq.~\eqref{e:QME1} with a Hamiltonian summing over qubits $\ell = 1$ and $2$,
\begin{equation}\label{eq:H}
  H = \sum_{\ell=1,2} \alpha_\ell\sigma_\ell^+ + \alpha_\ell^*\sigma_\ell^- + \Delta_\ell\sigma_\ell^+\sigma_\ell^-,
\end{equation}
where the Pauli raising and lowering operators are $\sigma^+ = (\sigma^-)^\dagger = \begin{pmatrix}0 & 0\\1 & 0\end{pmatrix}$. We define the operators $\sigma_1=\sigma\otimes I_{2 \times 2}$ and $\sigma_2=I_{2 \times 2}\otimes \sigma$ for a given Pauli operator $\sigma$. The type of Pauli operator is represented with a superscript
\begin{equation*}
  \sigma^{(x)} = \begin{pmatrix} 0 & 1\\ 1 & 0 \end{pmatrix}, \;
  \sigma^{(y)} = \begin{pmatrix}0 & \imath\\-\imath & 0\end{pmatrix}, \;
  \sigma^{(z)} = \begin{pmatrix}1 & 0\\0 & -1\end{pmatrix}.
\end{equation*}
The parameters $\alpha$ and $\Delta$ represent the qubit's Rabi frequency and the detuning of the qubit drive from the cavity, respectively. The qubit-cavity coupling via the cavity mode can then be described by the quantum jump operator
\begin{equation}\label{eq:V_c}
  V_c = s_1 \sigma_1^- + s_2 \sigma_2^-,
\end{equation}
with $s_{1,2}$ scalar. We emphasize that this operator represents \emph{collective} coupling between the qubits due to the two terms arising in $V_c$, and is distinctly separate from the single-qubit decay operators $V_{(1,r)} = \gamma_1^{(r)} \sigma_1^-$ and $V_{(2,r)} = \gamma_2^{(r)} \sigma_2^-$ that disturb the system, as do the single-qubit decoherence operators $V_{(1,\phi)} = \gamma_1^{(\phi)} \sigma_1^{(z)}$ and $V_{(2,\phi)} = \gamma_2^{(\phi)} \sigma_2^{(z)}$.

We work within the Bloch formalism~\cite{PhysRevA.93.063424,PhysRevA.81.062306,JPhysA37,neat_formula} that allows us to represent density matrices as vectors, in order to bring our system closer to the language of classical control. This formalism is derived from the classical technique of expanding $\rho$ with respect to a suitable orthonormal basis $\{\nu_n\}_{n=1}^{N^2}$, for the Hermitian operators on the Hilbert space $\H$ of dimension $N=4$. We can choose the basis such as $\nu_{N^2}=\tfrac{1}{N} \I$, where $\I$ is the identity operator in dimension $N^2$. Then defining $\vec{r}=(r_n)_{n=1}^{N^2}$ with $r_n = \Tr(\bm \sigma_n\rho)$, the basic operator $H$ and superoperator $\mathfrak{L}$ are mapped to $N^2 \times N^2$ real matrices as follows (defining $\{A,B\} = AB+BA$ as the anticommutator)
\begin{align*}
  (\A_H)_{mn}      &= \mathrm{Tr}(\imath H[\nu_m,\nu_n]), \\
  (\A_{V_k})_{mn}  &= \sum_k\gamma_k^2\mathrm{Tr}\left(V_k^\dag\nu_m V_k\nu_n  -\tfrac{1}{2}V^\dag_k V_k\{\nu_m,\nu_n\}\right).
\end{align*}
Defining $\A=A_H+\sum_k \A_{V_k}$ and a vector $\vec{c}$ with
\begin{equation}\label{eq:cvec}
  c_m = \frac{1}{N}\sum_k \gamma_k^2\Tr{\big ([V_k, V_k^\dagger]\nu_m \big )},
\end{equation}
we can define either an equation for the full state $\vec{r}$ that includes the pole at $0$ or an equation for the reduced state $\vec{r}_1$ that does not include the trivial dynamics of $\mbox{Tr}(\rho)$ and hence does not involve the pole at $0$,
\begin{equation}\label{eq:s_vec}
  \vec{\dot{r}} = \A\vec{r}, \quad \dot{\vec{r}}_1 = \A_{11}\vec{r}_1 + \vec{c},
\end{equation}
where $\A_{11}$ is a $(N^2-1)\times (N^2-1)$ matrix containing the first $N^2-1$ rows and columns of $\A$. Both forms equivalently describe the system dynamics and the second is readily solved to find the steady-state solution by setting the left-hand side to zero. One can also go between $\vec{r}$ and the density matrix $\rho$ by inverting the expansion. This is useful because, in what follows, we use the \emph{concurrence}~\cite{EntanglementConcurrence} at steady-state as a performance measure. The concurrence (which, as expected, is nonlinear in the state $\rho$) provides a measure of entanglement between two qubits, where $\tilde{C} = 1$ denotes a fully entangled quantum state (e.g., a Bell state):
\begin{equation}\label{eq:concurrence}
  \tilde{C}(\rho) = \max{(0, \lambda_1 - \lambda_2 - \lambda_3 - \lambda_4)},
\end{equation}
where the $\lambda_k$ are the eigenvalues (ranked from largest to smallest) of the matrix $R = \sqrt{\sqrt{\rho}\tilde{\rho}\sqrt{\rho}},$ and $\tilde{\rho}$ is given by
\begin{equation}\label{eq:concurrence_rhotilde}
  \tilde{\rho} = (\sigma^{(y)}\otimes\sigma^{(y)})\rho^*(\sigma^{(y)}\otimes\sigma^{(y)}),
\end{equation}
where $\rho^* = \rho^T$ because $\rho$ is Hermitian. In addition, we consider the fidelity $F$ of the steady state $\rho_\mathrm{ss}$ with respect to a target state $\rho_\mathrm{b}$. As our bare state (defined in detail in the next subsection) has unit purity (i.e. $\Tr{\rho_\mathrm{b}^2} = 1$), we define
\begin{equation}\label{eq:fidelity}
  F(\rho_\mathrm{ss}) = \Tr\{\rho_\mathrm{b}\rho_\mathrm{ss}\},
\end{equation}
which, if $\rho_\mathrm{b}$ is known, is linear in the steady state. Note that the purity is bounded between $1/N \leq \Tr{\rho^2} \leq 1$ for a state with Hilbert space dimension $N$.

\subsection{\label{subsec:param}Parameter values and units}

To illustrate the utility of the various methods considered in this manuscript, we consider the following scenario. Firstly, we will work with the following bare parameters, cf. Eq.~\eqref{eq:H}: $\alpha_1 = \alpha_2 = 1$, $\Delta_1 = -\Delta_2 = 0.1$, and $s_1 = s_2 = 1$, giving a concurrence of $C_\mathrm{bare} = 0.995$ for a bare state described by a density matrix $\rho_\mathrm{b}$. For systems of superconducting transmon qubits like those described in Ref.~\cite{minev_thesis}, these values are typical given that, for each parameter, a unity value corresponds to a frequency of $10$~MHz. In these systems, typical values for the $\gamma^{(r)}$s and $\gamma^{(\phi)}$ are about $0.001$, but we will consider here a wider range of decay and decoherence rates in an attempt to find the point at which system entanglement is lost completely, which can inform device design.

Furthermore, we consider the set of structured perturbations $S(\alpha_1, \alpha_2, \Delta_1, \Delta_2, s_1, s_2, \gamma^{(r)}_1, \gamma^{(r)}_2, \gamma^{(\phi)}_1, \gamma^{(\phi)}_2)$, similar to those considered in~\cite{robust_performance_open} (using the same notation as \cite{robust_performance_open}):
\begin{itemize}
  \item $S_2 = \{0,1,0,0,0,0,0,0,0,0\}$
  \item $S_4 = \{0,0,0,1,0,0,0,0,0,0\}$
  \item $S_5 = \{0,0,0,0,1,1,0,0,0,0\}$
  \item $S_7 = \{0,0,0,0,0,0,0,1,0,0\}$
  \item $S_9 = \{0,0,0,0,0,0,0,0,0,1\}$
  \item $S_{10}= \{0,0,0,0,1,-1,0,0,0,0\}$,
\end{itemize}
where the perturbations $S_5$ and $S_{10}$ change the collective coupling in a symmetric and antisymmetric way, respectively. However, due to the nonlinear dependence on the quantum jump operators in Eq.~\eqref{e:QME2}, the effects of these perturbations is not strictly additive when considered below and one has to be more careful when considering general perturbations to $s_1$ and $s_2$. We consider only positive perturbations on $S_7$ and $S_9$ since negative decay and decoherence make no sense physically.  With the exception of the coupled decay $S_5$ and $S_{10}$, all Hamiltonian perturbations act on qubit $2$; the results for perturbations on qubit $1$ are identical due to symmetry. 
In addition, we expect that, for this problem, all perturbations are symmetric about $s = 0$, where $s$ is the perturbation frequency.

Future work on the physics of this system can expand the methodology described here to perturbations on qubit 1 as well as the consideration of simultaneous perturbations on both qubits beyond those considered here.

\section{Results}

\begin{figure*}[t!]
\includegraphics[width=0.9\textwidth]{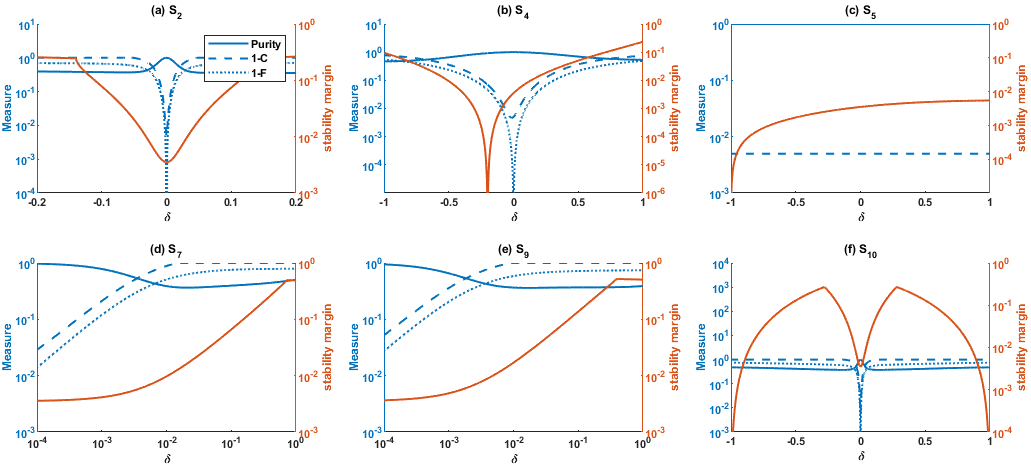}
\centering
\caption{State purity $\mathrm{Tr}(\rho_\mathrm{ss}^2)$ (left axis, blue solid), concurrence error $E_C=1-C$ (left axis, blue dashed), fidelity error $E_F = 1-F$ (left axis, blue dotted) relative to the unperturbed steady state and classically-inspired stability margin $G$ (right axis, orange solid) as a function of the perturbation $\delta$ on (a) $\alpha_2$ ($S_2$), (b) $\Delta_2$ ($S_4$), (c) symmetric perturbations on $s_1$ and $s_2$ ($S_5$), (d) $\gamma_2^{(r)}$ ($S_7$), (e) $\gamma_2^{(\phi)}$ ($S_9$), and (f) antisymmetric perturbations on $s_1$ and $s_2$ ($S_10$). Note that for all plots the x-axis is on a linear scale, except for (d) and (e) where the x-axis is logarithmic.}\label{fig:CFPsm}
\end{figure*}

\subsection{Structured perturbations of nonlinear entanglement and stability margin and linear fidelity}

In this section, we study structured perturbations of the system defined in Sec.~\ref{subsec:param} with an eye towards the concurrence as our performance measure. In particular, we consider the system $\A_k = \A + \delta_k \S_k$ for $k=2,4,5,7,9,10$ where $\delta_2\in [-0.2,0.2]$, $\delta_4\in [-1,1]\Delta_2$, 
$\delta_5 \in [-1,1]$, $\delta_7,\delta_9\in [0,1]$ to ensure physical constraints (e.g., non-negative decoherence rates) are satisfied.

Before diving into the classical control results, we study how the system purity, concurrence error $E_{\tilde{C}} = 1-\tilde{C}$, fidelity error $E_F = 1-F$, and the classically-inspired stability margin, defined as
\begin{equation}
  G = |\max_{\lambda_n(\A+\delta_k\S_k) \ne 0}\lambda_n(\A+\delta_k\S_k)|,
\end{equation}
where $\lambda_n(\A+\delta_k\S_k)$ are the eigenvalues of $\A+\delta_k\S_k$, change as we vary the parameters $\alpha_2$, $\Delta_2$, and $s_1 + s_2$. 
The results in Fig.~\ref{fig:CFPsm}(d) and (e) show that, when varying the decoherence and decay rates, there is a clear, traditionally inspired~\cite{Safonov_Laub_Hartmann}, trade-off between the classical stability margin and the concurrence in that lower concurrence implies a higher stability margin, in line with the observation that more decoherence and decay render the system less ``quantum''.  However, there is also some surprising behavior: one might expect the steady state of the system in the presence of decay to rapidly become trivial with both qubits in their ground state and no phase coherence between them as spontaneous emission is an incoherent process.  
Likewise, under decoherence, one would expect concurrence to rapidly decay as entanglement depends strongly on coherence between the two qubits.  However, substantial concurrence persists in the steady state up to decay rates of $\approx 0.01$, i.e., entanglement can persist even in a system under high degrees of decoherence and decay.

In general, we see that the system is slightly more sensitive to concurrence error than fidelity error. The lack of sensitivity in symmetric perturbation of $s_1 + s_2$ (with $s_1 = s_2$) is expected given that, for $\alpha_1 = \alpha_2 = \alpha$, $\Delta_1 = \Delta -\Delta_2$, $s_1 = s_2 = s_c/2$ and $\Delta_1 \ll \alpha$ (as satisfied roughly by our bare parameters), we can approximate $C \approx 2\alpha^2/(\Delta^2 + 2\alpha^2)$~\cite{motzoi}. This expression is independent of $s_c$, but if $s_c = 0$, the stability margin is zero and the concurrence error is $1$ due to the lack of any dissipative coupling (i.e., the system evolution is purely unitary).  However, any break in symmetry of the system causes the concurrence to drop rapidly.  The concurrence drops off rapidly as a function of $\alpha_2$ compared to perturbations on $\Delta_2$, implying that qubit entanglement is far more sensitive to the Rabi frequency of the drive than to its detuning; such analyses are extremely important when considering what real parameters one should make the effort to properly stabilize in an experimental system.
Anomalous behavior arises at $\Delta_2 = \Delta_1$ (i.e. $\delta_4 = -0.2)$, where the classical stability margin drops near zero, indicating that at least one pole is moving towards zero.  It is, however, impossible here to distinguish whether this is a single pole or whether the system is moving towards unitary behavior. However, where $s_1 = s_2 = 0$ (i.e. $\delta_5 = 1$), the Lindblad term coupling the qubits to the cavity is zero and thus the system evolves without any nonzero quantum jump operators; its evolution is governed only by the quantum Hamiltonian $H$ and thus the system behaves unitarily, although, interestingly, it maintains its concurrence up to (but not including) this point. This indicates that, as long as some non-zero coupling is present, the steady-state of the system has high concurrence.

\subsection{Structured perturbation of the error transfer matrix\label{subsec:struct}}

We now turn to the question of whether we can recover the same physics from classical structured perturbation analysis. After constructing the Bloch matrix $\A$, while considering only the unperturbed Hamiltonian and the lossy dual-qubit coupling term $V_c$, we compute the transfer matrix
\begin{equation}
  \T_{\vec{z},\vec{r}}^{(k)}(s,\delta) = (s\I-\A-\delta \S_k)^\#\delta \S_k
\end{equation}
from the unperturbed or nominal dynamics $\vec{r}$ to the error dynamics $\vec{z} = \vec{r}_\delta-\vec{r}$ via the perturbed dynamics $\vec{r}_\delta$ as introduced in~\cite{robust_performance_open}, 
where $\S_k$ is a structured perturbation of strength $\delta$ and $s$ is the Laplace variable. The Bloch matrix and its structured perturbation take the form
\begin{equation} \label{eq:AS}
  \A = \begin{bmatrix} \A_{11} & \A_{12} \\ 0 & 0 \end{bmatrix}, \quad
  \S = \begin{bmatrix} \S_{11} & \S_{12} \\ 0 & 0 \end{bmatrix},
\end{equation}
where $\A$ has size $N^2\times N^2$ and $\A_{12}$ has size $(N^2-1) \times 1$.
\begin{equation}
    \label{eq:hash_inverse}
    (s\I-\A)^\# := \begin{bmatrix}
    (s\I'-\A_{11})^{-1} & 0 \\ 0 & 0 \end{bmatrix}
\end{equation}
is the \emph{matrix \#-inverse} of $(s\I-\A)$ developed for such open quantum systems in Ref.~\cite{robust_performance_open}, with $\I'$ the $15\times 15$ identity matrix. It is then readily verified that $\widehat{\vec{z}}(s) = \T_{\vec{z},\vec{r}}(\delta,s) \widehat{\vec{r}}(s)$ with
\begin{align}\label{eq:bigT}
  \T_{\vec{z},\vec{r}}(s,\delta) = \begin{bmatrix}
    \bTheta_{11}(s,\delta)\delta\S_{11} & \bTheta_{11}(s,\delta)\delta\S_{12} \\
    0 & 0
  \end{bmatrix}
\end{align}
and $\bTheta_{11}(s,\delta)=(s\I'-\A_{11}-\delta \S_{11})^{-1}$. The zero eigenvalue associated with the constancy of the trace is removed while, as shown in~\cite{robust_performance_open}, most of the properties of $\bTheta(s,\delta)\delta \S$ for $s \ne 0$ are preserved; 
in particular $\lim_{s \to 0} \bTheta(s,\delta)\delta \S = \T_{\vec{z},\vec{r}}(0,\delta)$. Specifically, this shows that $\vec{z}_2\equiv 0$ and
\begin{equation}\label{eq:z1}
  \widehat{\vec{z}}_1 (s)
  = \bTheta_{11}(s) \delta \S_{11} \widehat{\vec{r}}_1
  + \bTheta_{11}(s) \delta \S_{12} \widehat{\vec{r}}_2.
\end{equation}
We can use the transfer function to study the effect of various structured perturbations and observe its behavior as a function of the perturbation strength $\delta$ and frequency $s = j\omega$.

\subsection{The classical transfer function as a bound for the distance between states}

\begin{figure*}[t!]
\centering
\includegraphics[width=0.9\textwidth]{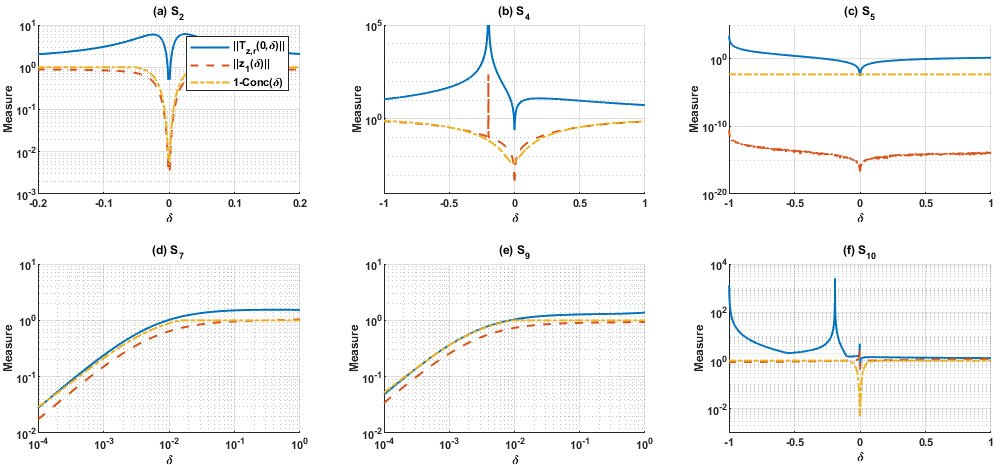}
\caption{Transfer function norm $\norm{\T_{\vec{z},\vec{r}}(0,\delta)}$ (blue, solid), distance bound $|\vec{z}_1(\delta)|$ (red, dashed), and concurrence error $E_C = 1-C$ (yellow, dotted) plotted as a function of $\delta$ for (a) $\alpha_2$ ($S_2$), (b) $\Delta_2$ ($S_4$), (c) symmetric perturbations on $s_1$ and $s_2$ ($S_5$), (d) $\gamma_2^{(r)}$ ($S_7$), (e) $\gamma_2^{(\phi)}$ ($S_9$), and (f) antisymmmetric perturbations on $s_1$ and $s_2$ ($S_{10}$).  Note that as in Fig.~\ref{fig:CFPsm} for all plots the x-axis is on a linear scale, except for (d) and (e) where the x-axis is logarithmic.}\label{fig:bounds}
\end{figure*}

To understand how the transfer function relates to a bound for the fidelity, recall that $\vec{r}$ follows the dynamics of the unperturbed system, $\dot{\vec{r}} = \A \vec{r}$ (cf.\ Eq.~\eqref{eq:s_vec}) with some initial condition $\vec{r}(0)$. Taking the Laplace transform gives $s\widehat{\vec{r}}(s) - \vec{r}(0) = \A \widehat{\vec{r}}(s)$ and $\widehat{\vec{r}}(s) = (s\I-\A)^{-1} \vec{r}(0)$ if $s\I-\A$ is invertible. Recalling $\T_{\vec{z},\vec{r}}(s,\delta) = (s\I-\A-\delta \S)^{-1} \delta \S$ gives
\begin{align*}
  \widehat{\vec{z}}(s) &= \T_{\vec{z},\vec{r}}(s,\delta)\widehat{\vec{r}}(s)\\
  &= (s\I-\A-\delta \S)^{-1} \delta \S (s\I-\A)^{-1} \vec{r}(0) \\
  &= [(s\I-\A-\delta \S)^{-1} -(s\I-\A)^{-1}]\vec{r}(0)
\end{align*}
where the equality in the third line comes from the matrix inversion lemma (as in Eq.~(13) in Ref.~\cite{robust_performance_open}). If the transfer function only has poles for $\Re(s)<0$ or $s=0$, then the final value theorem applies, and if $\A$ and $\A+\delta\S$ are invertible,
\begin{align}
  \lim_{t\to \infty} \vec{z}(t)
  &= \lim_{s\to 0} s \widehat{\vec{z}}(s) \nonumber\\
  &= \lim_{s \to 0}[-(\A+\delta \S)^{-1} + \A^{-1}]s\vec{r}(0) = 0,
\end{align}
i.e., the error asymptotically vanishes.

The situation is more complicated for the systems we are interested in due to the constants of motion. If the state vector $\vec{r}(t)$ follows the evolution of the unperturbed system then
\begin{align}
  \begin{bmatrix}
     \dot{\vec{r}}_1 \\ \dot{\vec{r}}_2
  \end{bmatrix}
  = \begin{bmatrix}
    \A_{11} & \A_{12} \\ 0 & 0
  \end{bmatrix}
  \begin{bmatrix}
    \vec{r}_1 \\ \vec{r}_2
  \end{bmatrix},
\end{align}
where $\vec{r}_2$ are the coordinates that are invariants of motion. Taking the Laplace transform
\begin{align}
  \begin{bmatrix}
    s \widehat{\vec{r}}_1 - \vec{r}_1(0)\\
    s \widehat{\vec{r}}_2 - \vec{r}_2(0)
  \end{bmatrix} =
  \begin{bmatrix}
    \A_{11} \widehat{\vec{r}}_1 + \A_{12}\widehat{\vec{r}_2}\\
    0
  \end{bmatrix}
\end{align}
gives $\widehat{\vec{r}}_2(s) = \vec{r}_2(0)/s$, and if $s\I'-\A_{11}$ is invertible,
\begin{equation}\label{eq:r1}
  \widehat{\vec{r}}_1(s) = (s\I'-\A_{11})^{-1} (\A_{12} \vec{r}_2(0)/s + \vec{r}_1(0)).
\end{equation}
Inserting Eq.~\eqref{eq:r1} into Eq.~\eqref{eq:z1} and using
\begin{align*}
  & (s\I'-\A_{11}-\delta\S_{11})^{-1}\delta \S_{11}\\
  &= [(s\I'-\A_{11}-\delta \S_{11})^{-1} -(s\I'-\A_{11})^{-1}] (s\I'-\A_{11})\\
  &= [\bTheta_{11}(s,\delta)-\bTheta_{11}(s,0)](s\I'-\A_{11}),
\end{align*}
which follows from the matrix inversion lemma, gives
\begin{equation*}
  \begin{split}
    \widehat{\vec{z}}_1
    =& [\bTheta_{11}(s,\delta)-\bTheta_{11}(s,0)] \, [\A_{12}  \vec{r}_2(0)/s+\vec{r}_1(0)]\\
     & + \bTheta_{11}(s,\delta) \delta \S_{12} \vec{r}_2(0)/s\\
    =& \bTheta_{11}(s,\delta)(\A_{12} +\delta\S_{12}) \vec{r}_2(0)/s  -\bTheta_{11}(s,0)\A_{12} \vec{r}_2(0)/s\\
     & + [\bTheta_{11}(s,\delta)-\bTheta_{11}(s,0)]\vec{r}_1(0)).
\end{split}
\end{equation*}

If $s\I'-\A_{11}$ and $s\I'-\A_{11}-\delta\S_{11}$ only have poles for $\Re(s)<0$ or $s=0$ then the final value theorem applies
\begin{align*}
    \lim_{t\to \infty} \vec{z}_1(t)
  &= \lim_{s\to 0} s \widehat{\vec{z}}_1(s) \nonumber\\
  &= [\bTheta_{11}(0,\delta)(\A_{12}+\delta\S_{12})-\bTheta_{11}(0,0)\A_{12}] \vec{r}_2(0).
\end{align*}

For a generic quantum system, there is usually one constant of motion, the trace of the density operator, and using a standard orthonormal basis, $\vec{r}_2(0)=1/N$, $\A_{12}\vec{r}_2(0)=\vec{c}$ and $\bTheta_{11}(0,0)\vec{c}$ is the steady-state $\vec{r}_{1,\rm ss}(0)$ of the unperturbed system and $\bTheta_{11}(0,\delta) (\A_{12}+\delta\S_{12})/N$ is the steady-state $\vec{r}_{1,\rm ss}(\delta)$ of the perturbed system. In particular, the steady-state depends only on the constants of motion and is independent of $\vec{r}_1(0)$, the dynamic part of the initial state.

The distance between the steady states of the perturbed and unperturbed system is bounded by the norm of the transfer function at $s=0$ and $\lim_{s\to 0} s\widehat{\vec{r}}(s)$.  If $\A_{11}$ is invertible then we can infer from Eq.~\eqref{eq:r1} and $\vec{\widehat{r}}_2(s) = (Ns)^{-1}$ that $\lim_{s\to0} s\widehat{\vec{r}}(s) =[-\A_{11}^{-1}\A_{12};1]/N=:\vec{d}$ and thus
\begin{equation}
\begin{split}
  \|\vec{r}_{\rm ss}(\delta) -\vec{r}_{\rm ss}(0) \|
  & = \lim_{s\to 0}\norm{\T_{\vec{z},\vec{r}}(s,\delta) s\widehat{\vec{r}}(s)}\\
  &\le \norm{\T_{\vec{z},\vec{r}}(0,\delta)}\| \vec{d}\|.
\end{split}
\end{equation}
If both states $\vec{r}_{\rm ss}(\delta)$ and $\vec{r}_{\rm ss}(0)$ are pure (which the latter is by definition), then $|\vec{r}| = 1$. This can be related to the fidelity error by noting that the fidelity can be written $F(\delta) = \vec{r}_{\rm ss}(\delta)^T\vec{r}_{\rm ss}(0)$. Thus, we have $\|\vec{r}_{\rm ss}(\delta) -\vec{r}_{\rm ss}(0) \|^2 = 2(1-F(\delta))$. However, this is generally not applicable, as the purity of the state is generally not unity, cf.\ Fig.~\ref{fig:CFPsm}.

\begin{figure}[t!]
\centerline{\includegraphics[width=0.8\columnwidth]{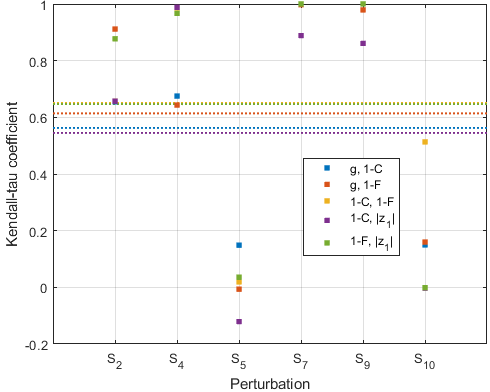}}
\caption{Concordances (via Kendall tau analysis) of variables in various pairs $a,b$ as shown in the legend: (blue) stability margin $G$, concurrence error $1-C$, (orange) $G$, fidelity error $1-F$, (yellow) $1-C$, $1-F$, (purple) $1-C$, $|\vec{z}_1|$ and (green) $1-F$, $|\vec{z}_1|$ in the $\delta$ ranges plotted in Figs.~\ref{fig:CFPsm} and~\ref{fig:bounds} with respect to the different structured perturbations considered in this manuscript, as shown on the x-axis. Each dashed line shows the mean value of the Kendall tau coefficient across all structured perturbations.}\label{fig:KT}
\end{figure}

To understand this in more detail, we plot $|\vec{z}_1(\delta)| = \|\vec{r}_{\rm ss}(\delta) -\vec{r}_{\rm ss}(0) \|$, $\norm{\T_{\vec{z},\vec{r}}(0,\delta)}$ and $1-C$ in Fig.~\ref{fig:bounds}. We see that, not unexpectedly, these bounds do not contain the concurrence, but there is some potential concordance. To understand the degree to which the bounds are concordant or discordant with the concurrence error, we ran Kendall tau analyses between the concurrence error $1-C$ and: the stability margin $G$, fidelity error $1-F$, and the $|\vec{z}_1|$ bound (three separate analyses for three separate pairs of variables). In addition, we looked at the concordance between $1-F$ and $G$, as well as $1-F$ and the $|\vec{z}_1|$ bound (an additional two analyses with two pairs of variables). The results are tabulated and plotted in Fig.~\ref{fig:KT}. We see that for all perturbations, except those affecting the qubit-cavity couplings ($S_5$ and $S_{10}$), there is general concordance ($\tau > 0.6$) between $1-C$ and the stability margin, as well as $1-C$ and the $|\vec{z}_1|$ bound; similar concordance can be seen between $1-F$ and these variables. Likewise, and somewhat unsurprisingly, we find general concordance between $1-C$ and $1-F$, that is, the concurrence error and fidelity error typically follow each other. However, for the qubit-cavity coupling perturbations, we find that concordance is minimal at best and the $|\vec{z}_1|$ bound is even discordant relative to the concurrence error. Note, however, that for the $S_5$ bound, the concurrence is completely flat for all $\delta\neq-1$, and the fidelity similarly shows little change as we scan $\delta$, and, away from a small region around $\delta = 0$, the same is true for $S_{10}$, although the former indicates that the system is completely robust to symmetric variations in $s_1$ and $s_2$ while the latter indicates a strong lack of robustness when this symmetry is broken, representing two very different scenarios. As such, while this bound can be a useful linear parameter that can monitor the nonlinear concurrence, this is not universal, and thus such concordance should be checked before, e.g., using the $|\vec{z}_1|$ norm as a performance measure in lieu of the concurrence, especially in cases where the concurrence or fidelity is particularly robust to a given perturbation. Thus, the general utility of this method is limited, which serves to further obviate the need for a theory of quantum control that works independently of classical robust control, even if some ideas can be gleaned from classical robust control.

\section{Discussion and Future Work}

The results shown in this manuscript demonstrate three main ideas: (a) When considering an open quantum system, where there is some decoherence, there is a quantum/classical tradeoff in that the classically-inspired stability margin increases as the ``quantumness'' of the system decreases (due to increased dissipation or dephasing). (b) When considering other structured perturbations (i.e. those not driving deleterious interactions with the environment), one cannot guarantee concordance between their size 
and the classically-inspired stability margin, especially when considering nonlinear performance indices like the concurrence. (c) Even though rigorous bounds on the fidelity error (but not the concurrence error, due to its nonlinearity) can be derived for pure states, these bounds fail in the case of any state impurity, and they provide little real physical insight. 

These statements should be couched, however, in the fact that we consider only one example system, and we make no claim that this is a rigorous proof that such methods would not apply better to other open quantum systems. The purpose of this manuscript is simply to try and glean physical insight from the application of classical control techniques to a given quantum system. Due to the relative lack of success in this area, we would like to highlight the need for a general quantum robust control theory that will allow for the derivation of real, useful bounds outside of the perturbative regime that apply to any general quantum system and do not require techniques like Monte-Carlo sampling, which can be computationally expensive, but are commonplace in the field due to the lack of a sensible alternative.

\end{document}